\RequirePackage[2022/05/31]{latexrelease}
\documentclass[conference, 10pt]{IEEEtran}
\IEEEoverridecommandlockouts

\usepackage{cite}
\usepackage{amsmath,amssymb,amsfonts}
\usepackage{algorithmic}
\usepackage{graphicx}
\usepackage{textcomp}
\usepackage{xcolor}
\usepackage{booktabs}

\usepackage{etoolbox}
\usepackage{nidanfloat}
\usepackage{flushend}
\usepackage{cite}
\usepackage{hyperref}
\usepackage[spaces,hyphens]{xurl}
\usepackage{caption}
\usepackage{subcaption}
\usepackage{tabularx}
\usepackage{amssymb}
\usepackage{pifont}
\usepackage{float}
\usepackage{tikz}
\usetikzlibrary{arrows.meta, positioning}

\usepackage{forest}

\newcommand\systemname{\textsc{StarveSpam}}

\begin{document}

\title{
\systemname{}: Mitigating Spam with Local Reputation in Permissionless Blockchains
\thanks{This work was funded by NWO/TKI grant BLOCK.2019.004.}
}

\author{Rowdy Chotkan,
        Bulat Nasrulin,
        J\'er\'emie Decouchant,
        Johan Pouwelse\\
        \{R.M.Chotkan-1, B.Nasrulin, J.Decouchant, J.A.Pouwelse\}@tudelft.nl \\
        \textit{Delft University of Technology}, The Netherlands}
        
\maketitle

\thispagestyle{plain}
\pagestyle{plain}

\begin{abstract}
Spam poses a growing threat to blockchain networks.
Adversaries can easily create multiple accounts to flood transaction pools, inflating fees and degrading service quality. Existing defenses against spam, such as fee markets and staking requirements, primarily rely on economic deterrence, which fails to distinguish between malicious and legitimate users and often exclude low-value but honest activity. To address these shortcomings, we present \systemname{}, a decentralized reputation-based protocol that mitigates spam by operating at the transaction relay layer. 
\systemname{} combines local behavior tracking, peer scoring, and adaptive rate-limiting to suppress abusive actors, without requiring global consensus, protocol changes, or trusted infrastructure.
We evaluate \systemname{} using real Ethereum data from a major NFT spam event and show that it outperforms existing fee-based and rule-based defenses, allowing each node to block over 95\% of spam while dropping just 3\% of honest traffic, and reducing the fraction of the network exposed to spam by 85\% compared to existing rule-based methods.
\systemname{} offers a scalable and deployable alternative to traditional spam defenses, paving the way toward more resilient and equitable blockchain infrastructure.
\end{abstract}

\begin{IEEEkeywords}
blockchain, spam mitigation, decentralized reputation, peer-to-peer networks, Sybil tolerance
\end{IEEEkeywords}

\section{Introduction}

\begin{figure*}[t]
\centering
\begin{forest}
for tree={
    grow'=south,
    draw,
    rounded corners,
    node options={align=center, font=\footnotesize, text width=2.8cm},
    edge={->},
    parent anchor=south,
    child anchor=north,
    l sep=8pt,
    s sep=4pt
}
[Blockchain Spam
    [Identity Abuse
        [Sybil Airdrops \\
        Faucet Spamming \\
        \textit{e.g., governance attacks}]
    ]
    [Computational Abuse
        [Opcode Abuse \\
        Execution Bombs \\
        \textit{e.g., Eth. DoS 2016}]
    ]
    [Storage Attacks
        [Dusting (UTXO bloat) \\
        Data Inscriptions \\
        \textit{e.g., Ordinals/BRC-20}]
    ]
    [Economic Exploits
        [Gas Bidding Wars \\
        Priority Sniping \\
        \textit{e.g., Ethereum mints}]
    ]
    [Transaction Flooding
        [Mempool Saturation \\
        Block Inclusion Abuse \\
        \textit{e.g., Solana NFT bots}]
    ]
]
\end{forest}
\caption{Taxonomy of blockchain spam attacks by resource exhaustion or unfair access.}
\label{fig:taxonomy}
\end{figure*}
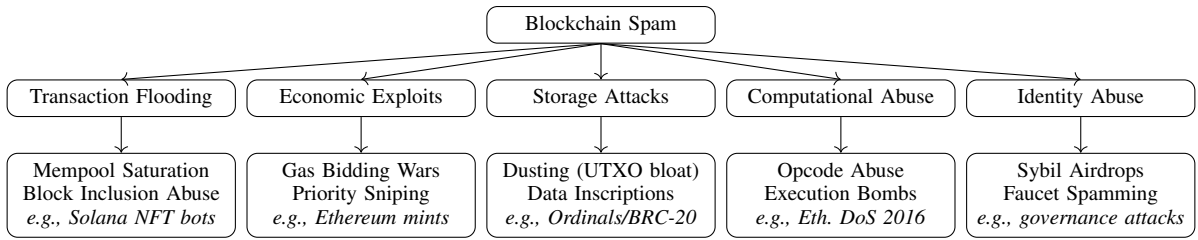

The Internet has been plagued by spam for nearly half a century. In 1978, an unsolicited advertisement for a DEC mainframe was sent to hundreds of ARPANET users, now considered the first recorded case of email spam~\cite{ferrara2019history,first_spam}. Nearly five decades later, spam at the various layers of digital systems continues to impact their performance. Despite substantial advancements in filtering techniques, platform-level moderation, and machine learning, spam remains a persistent and evolving threat~\cite{ahmed2022machine}. In 2024 alone, over 48\% of global email traffic was classified as spam~\cite{kaspersky_2024}. Centralized platforms, such as Gmail, Facebook, and Twitter, have developed powerful anti-spam tools but rely heavily on central control, content surveillance, and user profiling~\cite{ferrara2019history, meta_transparency}. In decentralized systems, however, spam remains a fundamental and unresolved challenge.

Because they do not benefit from centralized moderators or global coordination, decentralized systems struggle to contain spam. In particular, in blockchain networks, spam takes on new forms with broader implications. Unlike email spam, which primarily targets end-users, blockchain spam consumes shared global resources such as block space, network bandwidth, and node computation. Protocols such as Bitcoin, Ethereum, and Solana must defend against spam-like behavior that clogs mempools, inflates fees, and degrades network responsiveness.

In April 2022, Solana suffered a major outage when spam bots submitted over 4 million NFT mint transactions per second, saturating validator bandwidth and memory, and halting the chain for seven hours~\cite{solana2022incident,coindesk2022}. This event marked the beginning of a period of recurrent congestion. Subsequent incidents saw over 70\% of non-vote transactions fail, often due to spam bots flooding the mempool during NFT mints and speculative token launches~\cite{zheng2025does,blockworks_solana_spam}. In Bitcoin, spam campaigns have included `dust attacks'~\cite{binance2020dust}, where attackers send thousands of small outputs to wallets, bloating the UTXO set. In 2015, an attack filled blocks with junk outputs and consumed over 200\,BTC in fees to delay transactions~\cite{baqer2016stressing}. Ethereum has experienced similar disruptions, including a spam attack in 2016 that exploited opcodes to exhaust resources~\cite{ethereum2016dos}.

The dominant mitigation strategy in blockchain is economic deterrence. Protocols rely on various mechanisms to disincentivize abuse, such as fixed and dynamic minimum fees~\cite{saa2023iota}, fee auctions (e.g., EIP-1559~\cite{buterin2019eip1559}), and staking-based access control systems~\cite{buterin2013ethereum}.
While these approaches are practical against large-scale flooding, they suffer from two significant flaws. First, they fail to distinguish user intent: honest participants are penalized along with attackers during periods of congestion. Second, determined attackers can still outbid legitimate users, as evidenced by gas bidding wars during high-demand events such as NFT mints and airdrops~\cite{Jones_2022}.

These limitations arise from a deeper structural asymmetry in blockchain systems: creating a new identity is cheap, and costs are incurred only at execution time. Without persistent accountability or reputation, adversaries can cheaply generate new addresses or contracts to sustain flooding attacks. While economic deterrence offers a scalable defense, its side effects, particularly the exclusion of low-value but legitimate transactions, pose barriers to accessibility and fairness, especially for users operating at the network's margins. This points to a broader issue: spam in decentralized systems is not merely an economic nuisance, but a \emph{reputation problem}.

In traditional systems, reputation plays a central role in spam mitigation. Email servers use sender reputation and content-based heuristics (e.g., SPF, DKIM, and spam filters) to block abusive accounts~\cite{ahmed2022machine}. Social media platforms track behavioral signals (e.g., account age, follower networks, prior reports) to suppress low-quality or abusive actors~\cite{ferrara2019history, meta_transparency}. Even the Internet infrastructure relies on domain reputation and IP blacklists (e.g., DNSBLs~\cite{spamhaus}) to combat spam across networks~\cite{levine2010dns}. These defenses all depend on long-lived identity: misbehavior is remembered, and repeat offenders are penalized. In contrast, blockchains lack any notion of such continuity. There is no memory of past misbehavior, no account scoring, and no way to distinguish reliable senders from new or malicious ones. Each transaction is treated in isolation, allowing spammers to bypass filtering simply by rotating addresses. This absence of reputation memory makes spam particularly hard to contain in decentralized environments.

To address these issues, we present \systemname{}, a decentralized, reputation-based spam mitigation mechanism. Our approach combines long-lived identities with a Sybil-tolerant reputation system, allowing each node to locally assess peer behavior over time and respond accordingly. Rather than filtering individual transactions or enforcing global bans, our mechanism applies subjective and adaptive rate-limiting, preserving decentralization while curbing abuse. \systemname{} offers a novel direction for decentralized spam resistance, moving beyond fees and toward accountability without centralization.

As a summary, this work makes the following contributions:

\noindent $\bullet$ We propose a taxonomy of transaction-layer spam attacks, organizing common abuse patterns by the resources they exhaust or mechanisms they exploit.

\noindent $\bullet$ We design and prototype \systemname{}, a decentralized spam mitigation protocol that operates at the transaction relay layer using local reputation scores and adaptive rate-limiting.

\noindent $\bullet$ We evaluate \systemname{} using real-world Ethereum data from a large-scale NFT spam event, as well as synthetic scenarios with varying traffic and attacker behavior.

\section{Problem Description}
Spam in blockchain systems refers to the injection of transactions that serve little or no economic purpose but consume shared network resources such as bandwidth, block space, and validator computational resources. These transactions are often submitted in high volumes to degrade network responsiveness, drive up fees, or crowd out legitimate users. While spam has been extensively studied in contexts such as email~\cite{ferrara2019history} and peer-to-peer overlays~\cite{walsh2005fighting,zhai2009spamresist}, it poses unique challenges in permissionless blockchain networks. In contrast to permissioned systems, where participants can be vetted or rate-limited at the network level, permissionless protocols, such as Ethereum, Solana, and Bitcoin, must process all syntactically valid, fee-paying transactions, regardless of the sender's identity or behavior. This makes spam not only a nuisance but a recurring attack vector for denial-of-service, fee manipulation, and state bloat.

\subsection{Taxonomy of Blockchain Spam}

Spam in blockchain systems can be classified along several dimensions, including attack surface, exploited resource, and adversarial intent. Prior taxonomies have examined blockchain attacks broadly, ranging from consensus faults to application-layer exploits (e.g., \cite{chen2022survey,alkhalifah2020taxonomy,lubes2023tree}), emphasizing the importance of organizing threats by their system impact. Building on this perspective, we introduce a new taxonomy of spam-specific attacks, grounded in real-world observations. We manually classified attack types by reviewing post-mortems and mapping them to system-level impact (e.g., fee inflation, state bloat, or validator DoS). Fig.~\ref{fig:taxonomy} categorizes these attacks by the primary resource they exhaust or the mechanism they exploit:

\begin{itemize}
\item \textbf{Transaction Flooding.} Attackers flood the network with transactions to saturate mempools or fill blocks, delaying honest activity and triggering fee spikes. For example, during Solana's 2022 NFT minting event, bots submitted over 4 million transactions per second, causing a multi-hour outage~\cite{solana2022incident,coindesk2022}.

\item \textbf{Economic Exploits.} Spam can manipulate fee mechanisms (e.g., gas wars or priority sniping~\cite{Jones_2022}) to crowd out honest users, especially during high-demand events, where well-funded spammers outbid others and undermine fairness.

\item \textbf{Storage Attacks.} Techniques such as dust attacks, NFT airdrop spam, and data inscriptions (e.g., Ordinals or BRC-20~\cite{coinmarketcap2023ordinals}) bloat the state or UTXO set, increasing long-term storage overhead for nodes.

\item \textbf{Computational Abuse.} Spam can also target validator resources, as in Ethereum's 2016 DoS attack, which exploited underpriced opcodes to exhaust CPU and memory resources~\cite{ethereum2016dos}.

\item \textbf{Identity-Based Exploits.} Spammers leverage disposable pseudonyms to farm airdrops, drain faucet funds, and influence votes~\cite{gitcoinsybil}. These Sybil-style attacks degrade incentive alignment in user-driven protocols.
\end{itemize}

While diverse in mechanism and impact, these spam attacks exploit a common structural weakness: the lack of long-term accountability in permissionless systems.

\subsection{Why Spam Persists}

The persistence of spam in decentralized networks stems from two root causes: 

\begin{itemize}
\item \textbf{Identities are cheap and disposable}: Participants can generate new addresses at no cost. There is no inherent notion of continuity, accountability, or history tied to a participant.

\item \textbf{Protocols are behavior-agnostic}: Block producers accept any valid transaction that pays a fee, treating spammers and legitimate users alike.
\end{itemize}

At their core, most blockchain protocols are stateless with respect to participant behavior. This neutrality may uphold decentralization, but it also enables attackers to repeat abuse without consequence.
These properties make blockchains particularly vulnerable to Sybil attacks, as adversaries can flood the system under many pseudonyms without consequences.

\subsubsection*{Limitations of Economic Defenses}

Current anti-spam mechanisms primarily rely on economic deterrence: minimum fees, dynamic fee markets (e.g., EIP-1559~\cite{buterin2019eip1559}), and stake-based access control. These tools increase the cost of spamming but suffer from two key limitations:

\begin{itemize}
\item \textbf{Collateral damage.} Fee pressure penalizes honest and malicious users alike, without regard for intent or value.

\item \textbf{Insufficient deterrence.} Motivated attackers can outbid others, especially when financial rewards (e.g., airdrops or MEV~\cite{nasrulin2023accountable}) outweigh the costs.
\end{itemize}

These defenses fail not because fees are ineffective, but because they ignore identity and behavioral history. Without the ability to distinguish trustworthy behavior from manipulation, protocols remain vulnerable to well-funded adversaries.

\subsubsection*{The Need for Behavioral Accountability}

These challenges highlight a fundamental gap: decentralized systems currently lack mechanisms for associating actions with persistent identity or memory. Without some form of behavioral accountability, protocols remain vulnerable to spam by rational or malicious actors who exploit protocol neutrality. To address this, we seek a decentralized approach that allows networks to retain, distinguish, and adapt to participant behavior, without compromising decentralization or permissionlessness.

\newcolumntype{C}{>{\centering\arraybackslash}X}

\begin{table*}[t]
\centering
\caption{Comparison of spam mitigation approaches in blockchain systems}
\label{tab:spam_comparison}
\begin{tabularx}{\textwidth}{lCCCCC}
\toprule
\textbf{Approach} & \textbf{Decentralized} & \textbf{Sybil-Tolerant} & \textbf{Behavior-Aware} & \textbf{Tx-Level} & \textbf{Protocol Changes} \\
\midrule
Fee Thresholds (EIP-1559, Solana) & \checkmark & -- & -- & \checkmark & -- \\
Static Heuristics (e.g., BanMan) & \checkmark & -- & $\sim$ & \checkmark & \checkmark \\
Reputation Filters (Zhang et al.) & \checkmark & $\sim$ & \checkmark & \checkmark & -- \\
Identity Systems (BrightID, Gitcoin) & $\sim$ & \checkmark & -- & -- & -- \\
\textbf{\systemname{} (Ours)} & \checkmark & \checkmark & \checkmark & \checkmark & -- \\
\bottomrule
\end{tabularx}
\end{table*}

\section{Related Work}

Spam resistance in decentralized networks has been approached from several angles, including fee mechanisms, propagation heuristics, reputation systems, and cryptographic rate-limiting. We discuss these mechanisms in the following.

\subsection{Economic Spam Deterrents}

Most blockchains rely on cost barriers to discourage spam. Bitcoin enforces minimum relay fees and dust limits~\cite{nakamoto2008bitcoin}, while Ethereum introduced EIP-1559 to dynamically adjust base fees under congestion~\cite{buterin2019eip1559}. IOTA requires Proof-of-Work for transaction submission~\cite{saa2023iota}, and Solana supports priority fees to bid for inclusion. EOS and Internet Computer (ICP) tie transaction capacity to staked system resources (e.g., CPU, RAM)~\cite{eoswhitepaper,dfinity2022internet}.

These mechanisms make large-scale spam costly, but do not differentiate intent or reputation. During high-demand periods (e.g., NFT mints), honest users may still be priced out by well-funded attackers~\cite{Jones_2022}. Moreover, spam campaigns, such as Bitcoin's 2015 stress test, demonstrated that adversaries may willingly spend hundreds of BTC to degrade service~\cite{baqer2016stressing}.

\subsection{Protocol-Level Heuristics}

Several blockchain clients use implementation-specific spam filtering. Bitcoin's \textit{BanMan} module tracks misbehaving peers (e.g., invalid blocks or headers) and discourages them by omitting them from peer discovery~\cite{bitcoin_banman}. However, bans are local, temporary, and not persistent across restarts. Ethereum has used opcode repricing to mitigate underpriced computational attacks such as the 2016 DoS campaign~\cite{ethereum2016dos}, and Solana is exploring localized fee markets per smart contract~\cite{angeris2024multidimensional}.

These defenses are often reactive and fragile. They require tuning, depend on protocol changes, and do not generalize across ecosystems.

\subsection{Reputation and Identity Mechanisms}

Reputation systems have been deployed in peer-to-peer overlays such as Credence~\cite{walsh2006experience}, EigenTrust~\cite{kamvar2003eigentrust}, and~ Merit\-Rank~\cite{nasrulin2022meritrank} to assess peer quality. Zhang \textit{et al.}~\cite{zhang2020preventing} propose a decentralized reputation mechanism for Bitcoin that rates peers based on the validity and usefulness of forwarded transactions. These systems typically assume strong identities and consistent peer interaction, which may not hold in all blockchain environments. Unlike this work, we explicitly embrace the assumption of persistent pseudonyms and propose a Sybil-tolerant design that uses local and behavior-based reputation for spam mitigation.

Projects such as BrightID, Gitcoin Passport, and Proof of Humanity\footnote{\url{https://brightid.org}; \url{https://gitcoin.co/passport}; \url{https://proofofhumanity.id}} attempt to create Sybil-resistant identities, but are typically deployed at the application layer rather than for transaction filtering. Spam resistance through identity in public blockchains remains underexplored.

\subsection{Privacy-Preserving and Challenge-Response Defenses}

Unlinkability and challenge-response have been used to rate-limit abuse without full identity. Inselvini \textit{et al.}\cite{inselvini2021spam} propose using ZK-SNARKs for anonymous, spam-resistant content sharing with centralized or federated gatekeepers. CAPTCHA and challenge-response methods like XEP-0377\cite{whited2016xep0377} allow user-driven blocking and spam reporting in XMPP-based systems. These approaches rely on user friction or central authorities and are not well-suited to low-latency, decentralized transaction propagation.

While prior work covers fee tuning, heuristics, and identity systems, each falls short in one or more dimensions, such as Sybil tolerance or behavioral awareness. Table~\ref{tab:spam_comparison} summarizes these trade-offs. In contrast, our approach combines long-lived pseudonyms with local, behavior-based reputation in a fully decentralized setting. It applies adaptive rate-limiting without global state, identity uniqueness, or centralized enforcement.

\begin{figure*}[t]
\centering
\begin{tikzpicture}[
  node distance=1.5cm and 0.8cm,
  every node/.style={draw, minimum height=1.1cm, minimum width=2.5cm, align=center, font=\small},
  arrow/.style={-{Stealth}, thick},
  feedback/.style={->, thick, dashed}
]

\node (a)   [draw] {Ingestion};
\node (b)   [right=of a]    {Behavior\\Monitor};
\node (c)   [right=of b]    {Reputation\\Manager};
\node (d)   [right=of c]    {Rate\\Controller};
\node (e)   [right=of d]    {Gossip\\Filter};

\draw[arrow] (a) -- (b);
\draw[arrow] (b) -- (c);
\draw[arrow] (c) -- (d);
\draw[arrow] (d) -- (e);

\path (e.south) -- ++(0,-0.5) coordinate (mid1)
                -- ++(-3.2,0) coordinate (mid2);

\draw[feedback] (e.south) |- (mid2) -| (c.south);

\node[below=0cm of mid2, font=\footnotesize, align=center, draw=none]
  {Propagation decision affects \\ future reputation};

\end{tikzpicture}
\caption{Overview of \systemname{}'s transaction pipeline.}
\label{fig:system-design}
\end{figure*}
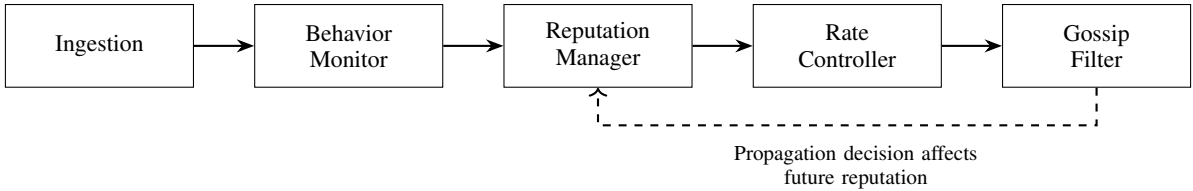

\section{System Design} \label{sec:system_design}
We design \systemname{} for permissionless blockchain networks, where nodes communicate over a peer-to-peer overlay and autonomously decide which transactions to accept, relay, or drop. Identities are cheap and ephemeral: nodes are identified by public keys and may generate pseudonyms at negligible cost. The network lacks central coordinators, shared state, or global reputation; each node relies solely on its own observations to assess peer behavior.

In this setting, we assume that adversaries may control many Sybil identities and generate large volumes of low-utility transactions. At the same time, honest nodes behave consistently and can build local trust over time. \systemname{} exploits this asymmetry: it enables nodes to classify transactions based on behavioral signals, assign local reputation scores to senders, and apply adaptive rate-limiting, all without protocol changes or global coordination.
\systemname{} operates entirely at the transaction relay layer. As illustrated in Fig.~\ref{fig:system-design}, it consists of modular components that allow each node to assess incoming transactions independently and degrade service for abusive senders, while preserving throughput and fairness for legitimate participants.

\subsection{Design Rationale}
\systemname{} is structured as a pipeline of modular components, each serving a distinct role in the local spam mitigation process. The pipeline reflects a logical dependency: transaction ingestion must precede behavioral classification, which in turn feeds reputation scoring, rate control, and gossip filtering. This modular design allows each node to make independent relay decisions while retaining the ability to adapt and tune specific components. We selected these components based on three criteria: (i) their feasibility in real-world node implementations; (ii) their alignment with observed spam behaviors; and (iii) their synergy in preserving decentralization while enabling behavioral accountability.

\subsection{Transaction Ingestion}

When a transaction is received from a peer, it first passes through the transaction ingestion layer, which extracts relevant metadata and routes it to downstream modules. Specifically, the ingestion layer parses the sender's public key (from the transaction signature), the gas price or fee information, the calldata size and content hash, and the touched contract addresses or UTXOs (depending on the blockchain).

In permissionless networks like Ethereum or Solana, senders are identified by public keys, which are cheap to generate. While this prevents strong identity, we assume that most spam campaigns operate under persistent pseudonyms (e.g., fixed wallets or contract deployers) due to cost structures (staking, airdrop eligibility, access to private keys). This assumption is consistent with prior work that shows that long-running spam campaigns reuse addresses for operational or strategic reasons~\cite{baqer2016stressing}.

The ingestion layer does not perform any filtering itself, but prepares transactions for classification by computing per-transaction features used by the behavior monitor. It also maintains a rolling history of recent transactions from each sender to support local temporal analysis (e.g., rate tracking and inclusion ratios). The history is stored as a fixed-size queue of the last $N$ transactions or covering a time window of $\Delta T$ seconds, whichever is longer. In our prototype (see~\autoref{sec:evaluation}), we use $N=20$ and $\Delta T=60$.

\subsection{Behavior Monitoring}

The behavior monitor evaluates incoming transactions using a set of interpretable, rule-based heuristics. Its purpose is to detect abuse patterns, such as flooding, resource exhaustion, or data duplication, without relying on centralized labeling or black-box classifiers. Inspired by documented spam campaigns across Ethereum, Solana, and Bitcoin, we design the monitor around five transaction-level features commonly associated with abusive behavior:

\begin{itemize}
    \item \textbf{Low gas price or fee.} The transaction offers a gas price below the 25th percentile of recent activity, reflecting low-fee spam tactics used in attacks like Bitcoin's 2015 stress tests and Ethereum's gas wars~\cite{baqer2016stressing, lopp2021history}.
    
    \item \textbf{Redundant calldata.} The calldata or payload matches a recent transaction from the same sender, capturing duplication patterns seen in NFT mint spam and contract call floods~\cite{zheng2025does}.
    
    \item \textbf{High revert rate.} The sender shows a high rate of reverted transactions (e.g., $4$ out of the last $10$), indicating abusive behavior. Revert-heavy spam is common in DeFi bot floods and failed arbitrage attacks, especially on Solana~\cite{zhang2020preventing, blockworks_solana_spam}.
    
    \item \textbf{Burst frequency.} The sender has submitted more than $k$ transactions in a short time window (e.g., $5$ in $3$ seconds), a pattern typical of denial-of-service and congestion attacks across major blockchains~\cite{baqer2016stressing, coindesk2022}.
    
    \item \textbf{Non-inclusion penalty.} A large share of the sender's recent transactions remain unconfirmed after $\beta$ blocks (e.g., $7$ of $10$ pending after $20$ blocks), indicating low economic utility. This heuristic captures low-impact activity that bloats mempools without affecting state~\cite{zhang2020preventing,nasrulin2023sustainable}.
\end{itemize}

Each rule is lightweight and implementable using local node observations. Nodes may assign weights to each condition to compute a spam score or apply simple logic (e.g., flag as spam if $\geq 2$ conditions are met, as applied in our experiments). The classification result is then forwarded to the reputation manager to update the sender's long-term score.

This rule-based approach strikes a balance between robustness and explainability. It avoids the operational overhead of machine learning, while remaining adaptable; thresholds can be tuned, and new rules can be added as spam tactics evolve. Each feature is grounded in real-world attack patterns, ensuring both defensibility and empirical relevance.

\subsection{Reputation Management}
The reputation manager maintains a local, persistent score $r_i \in [0, 1]$ for each sender, based on classifications from the behavior monitor. This score serves as a proxy for trustworthiness, informing transaction admission and propagation decisions. In \systemname{}, reputation is strictly local and subjective: each node scores addresses independently, using only its own observations.

Each sender's score is updated using an exponentially weighted moving average (EWMA) of past classification outcomes, balancing responsiveness to recent behavior with resistance to noise. Transactions labeled as spam decrement the score, while those classified as benign increment it. The update function includes:

\begin{itemize}
    \item \textbf{Score decay.} To allow forgiveness over time, scores gradually return to neutral if no further spam is observed.

    \item \textbf{Recency bias.} Recent transactions are weighted more heavily than older ones, enabling the system to respond quickly to changes in behavior.

    \item \textbf{Capping and clipping.} Scores are bounded within a fixed range (i.e., $[0, 1]$) to simplify downstream rate control and filtering logic.
\end{itemize}

This design is inspired by reputation systems in peer-to-peer overlays (e.g., EigenTrust~\cite{kamvar2003eigentrust}, Credence~\cite{walsh2006experience}, Merit\-Rank~\cite{nasrulin2022meritrank}), and adapted for pseudonymous blockchain contexts. Since identities are derived from public keys and can be easily regenerated, \systemname{} is designed to be Sybil-tolerant:

\begin{itemize}
    \item Fresh identities with little history begin with a neutral reputation and are conservatively throttled until they demonstrate reliability.
    \item Rapid identity rotation provides no advantage, as spammers cannot accumulate reputation across short-lived addresses.
    \item Long-lived, honest participants gradually build a stronger reputation, enabling higher throughput and improved responsiveness.
\end{itemize}

This approach punishes repeat abusers while allowing recovery and resilience to misclassification. It avoids the brittleness of centralized blocklists and ensures that honest users, especially those with low activity or intermittent connectivity, are not permanently excluded. However, because reputation builds over time, new or infrequent participants may be conservatively throttled at first—a necessary trade-off to limit Sybil abuse. Nodes can tune initial scores or thresholds to better balance openness and caution.

\subsection{Rate Control and Admission Policy}
Once a sender's reputation score is updated, the rate controller determines how to handle subsequent transactions. This module enforces adaptive, local rate-limiting, prioritizing resources such as mempool space, bandwidth, and validation cycles for reputable participants, while progressively throttling low-reputation identities. We partition reputation scores into three regions to guide rate control decisions. For reference, in our evaluation (see~\autoref{sec:evaluation}) nodes with scores above $\tau_{\text{high}} = 0.8$ are considered high-reputation, those below $\tau_{\text{low}} = 0.2$ are low-reputation, and scores in between are treated as moderate-reputation. Each incoming transaction is processed based on the sender's current reputation $r_i$:

\begin{itemize}
    \item \textbf{High-reputation}---Immediate mempool admission; applies when $r_i \geq \tau_{\text{high}}$.
    \item \textbf{Moderate-reputation}---Subject to queuing or delay, especially under congestion; applies when $\tau_{\text{low}} \leq r_i < \tau_{\text{high}}$.
    \item \textbf{Low-reputation}---Deprioritized or dropped entirely under load, effectively starved of access until reputation improves; applies when $r_i < \tau_{\text{low}}$.
\end{itemize}

This design acts as a soft admission gate: it does not impose rigid bans, but degrades service quality in proportion to observed behavior. Unlike fixed fee thresholds or static opcode bans, rate control offers gradual, feedback-driven enforcement that balances fairness with responsiveness. The controller also adapts to system conditions. Under normal load, nodes may accept more low-reputation traffic to preserve openness. Under stress, thresholds tighten to prioritize trusted flows. This elasticity mirrors fee market dynamics but is guided by reputation rather than gas price. 

By mediating between the behavior monitor and the mempool, the rate controller ensures that abusive identities cannot monopolize resources, even when they adhere to fee policies. The rate controller is the mechanism through which reputation directly shapes transaction acceptance.

\subsection{Gossip Filtering}
In addition to controlling local mempool admission, \systemname{} influences how transactions are propagated across the network. The gossip filter leverages sender reputation to suppress the spread of spam transactions, reducing their visibility and limiting the chance they will reach validators or block producers via other peers. In most blockchain P2P layers (e.g., Ethereum's devp2p or Solana's Turbine), transactions are gossiped opportunistically to random or stake-weighted peers. By default, nodes forward transactions indiscriminately, consuming unnecessary bandwidth and memory during spam events. \systemname{} modifies this behavior by broadcasting transactions from high-reputation senders without restriction, while treating low-reputation traffic more conservatively: such transactions may be forwarded with reduced priority, sampled at a lower rate, or dropped under congestion or buffer pressure.

This mechanism ensures that even if one permissive peer accepts a spam transaction, it is unlikely to propagate widely, effectively containing its blast radius. Because reputation is local, each node may make different forwarding decisions, creating a form of subjective resistance that is hard to game. Spammers must rebuild trust independently with every peer they contact. This is particularly valuable in settings where spam is economically rational (e.g., Solana bots or arbitrage attacks), as it introduces non-fee-based pressure: spammers may continue paying for bandwidth, but their reach is throttled by network-level filtering. Additionally, the gossip filter acts as a natural reputation sink, progressively isolating abusive addresses and limiting their ability to amplify attacks through peer propagation.

\subsection{Deployment and Integration}
\systemname{} is modular, decentralized, and incrementally deployable. It operates entirely at the transaction relay layer and requires no changes to consensus, transaction formats, or global coordination. As such, it can be integrated into a wide range of blockchain clients and overlays, without disrupting core protocol functionality (e.g., Ethereum, Bitcoin, and L2 systems). Nodes run it independently as a local pre-processing module that filters, scores, and regulates incoming transactions. Since all components (classification, reputation tracking, rate limiting, gossip filtering) rely solely on local observations and subjective policies, there is no need to trust external inputs or converge on a global view. \systemname{} is therefore well-suited for permissionless, pseudonymous networks.

The system can be integrated at various points in the stack: full nodes use it to reduce mempool and P2P load, validators or sequencers apply reputation-aware admission to protect block production, light clients avoid acting as spam amplifiers, and rollups filter spam before costly settlement operations.

Rate control is local and modular. Nodes can tune parameters such as thresholds, decay rates, and queue sizes based on hardware constraints, risk tolerance, or network role, thereby calibrating trade-offs between openness and abuse resistance. Its modular architecture supports upgrades to detection logic or reputation models without coordination or hard forks. While the current design uses rule-based heuristics, the framework is extensible. Nodes may incorporate local machine learning models or integrate signals from off-chain reputation and identity systems. The core rate-control and reputation logic remains compatible with such enhancements.

\section{Experimental Evaluation} \label{sec:evaluation}
We evaluate \systemname{} through three experiments, each targeting a key aspect of decentralized spam mitigation. First, we replay a real-world spam event to assess filtering under short-term load. Second, we simulate a large-scale network to observe how local reputations evolve for honest, malicious, and oscillating nodes. Third, we model peer-to-peer gossip to evaluate how local filtering limits network-wide spam propagation.

\newcolumntype{L}{>{\raggedright\arraybackslash}X}
\begin{table}[b]
\centering
\caption{Transaction fields used in the dataset}
\label{tab:tx_fields}
\begin{tabularx}{\linewidth}{@{}lL@{}}
\toprule
\textbf{Field} & \textbf{Description} \\
\midrule
\texttt{timestamp} & The Unix timestamp of the transaction. \\
\texttt{from\_address} & Ethereum address that initiated the transaction. \\
\texttt{calldata} & Encoded input data for contract execution. \\
\texttt{gas\_price} & Gas price offered by the sende, in Gwei. \\
\texttt{receipt\_status} & Execution status (1 = success, 0 = revert). \\
\texttt{receipt\_gas\_used} & Amount of gas consumed during execution. \\
\bottomrule
\end{tabularx}
\end{table}

\subsection{Setup and Dataset} \label{sec:experiment1}
All experiments run in a custom discrete-event simulator built with SimPy, where each node independently applies local admission policies. We use both real-world Ethereum data and synthetic traces, measuring both per-node and network-wide outcomes under varied traffic and behavioral scenarios.

Our first and third experiments use a dataset of $50{,}000$ Ethereum transaction data from the Otherside NFT mint (April 30, 2022), an event that caused extreme congestion and over \$150 million in gas fees, with many transactions failing due to mempool flooding~\cite{vice_otherside, guardian_otherside}. The dataset comprises all public transactions during the event window, sourced from the Ethereum network, and covers the fields listed in Tab.~\ref{tab:tx_fields}. Our dataset includes only transactions that were eventually confirmed on-chain, omitting those dropped or delayed indefinitely. While this excludes some aggressive spam attempts, we capture a broad range of low-utility and reverted transactions representative of real-world abuse. The second experiment uses synthetic traces that simulate up to $100$ nodes, with scheduled patterns such as diurnal activity, spam bursts, and behavioral shifts.

To assign ground-truth labels, we apply a multi-heuristic classifier based on established indicators. A transaction is labeled as spam if it satisfies at least two of five such heuristics (detailed in Tab.~\ref{tab:spam_rules}). The classifier operates offline with fixed thresholds and relies solely on general, transaction-level features. While some features overlap with those used by \systemname{}'s behavior monitor, the labeling process is entirely decoupled, having no access to its internal logic. The classifier uses static rules to assign evaluation labels, whereas \systemname{} updates peer reputation over time and applies adaptive rate control. This separation avoids circularity and ensures that the system is not tuned to match the labeling strategy. 
Approximately $62\%$ of transactions are flagged as spam, consistent with post-mortems that report widespread failure and abuse~\cite{castor_otherside}.

\begin{table*}[b]
\centering
\caption{Heuristics Used in Rule-Based Spam Classification. Each rule is derived from documented attack patterns and designed for local, explainable evaluation.}
\label{tab:spam_rules}
\resizebox{\textwidth}{!}{%
\begin{tabular}{@{}p{3cm}p{6.2cm}p{7.2cm}@{}}
\toprule
\textbf{Feature} & \textbf{Description} & \textbf{Rationale} \\ \midrule

\textbf{Duplicate calldata} &
Transaction shares its calldata with one or more others in the trace. &
Large-scale spam campaigns often submit identical or near-identical transactions in bulk, aiming to exploit timing advantages during NFT drops or airdrop eligibility checks. Repetition is a strong signal of scripted behavior. \\

\textbf{Reverted execution} &
Transaction fails to execute and reverts (receipt status = 0). &
High revert rates indicate speculative, zero-signal activity (e.g., failing arbitrage attempts). These transactions consume compute and gas while producing no state changes, degrading system utility. \\

\textbf{Low fee} &
Gas price is below the 10th percentile of all transactions in the dataset. &
Spam bursts are frequently issued at minimum viable cost. Abusers exploit mispriced fee periods or attempt to sneak spam through low-priority lanes, especially during periods of low congestion. \\

\textbf{Low complexity} &
Gas used by the transaction falls below the 10th percentile of the trace. &
Simple, no-op, or filler transactions are characteristic of congestion attacks and spam waves. Such transactions provide little or no economic utility but burden bandwidth and processing. \\

\textbf{Burst activity} &
Sender issues multiple transactions in rapid succession within a few seconds. &
Spammers often launch transactions in bursts to saturate mempools or maximize the chance of inclusion. This is common in denial-of-service scenarios and competitive bidding situations. \\

\bottomrule
\end{tabular}%
}
\end{table*}

\subsection{Filtering Effectiveness under Spam Load}
In our first experiment, we measure how effectively \systemname{} filters spam while preserving honest participation during a high-volume, adversarial workload. We replay the labeled transaction trace from the Otherside NFT mint and compare \systemname{} against five baselines:

\begin{itemize}
    \item \textbf{Naive}: accepts all transactions unconditionally.
    \item \textbf{Fee filter}: drops transactions below the 10th percentile of gas price.
    \item \textbf{BanMan}: penalizes peers based on low-fee and reverted transactions.
    \item \textbf{EIP-1559}: drops transactions below a dynamic base fee (modeled as the 25th percentile gas price over a rolling window).
    \item \textbf{SIMD-110}: simulates per-account congestion penalties inspired by Solana's proposed write-lock markets.
\end{itemize}

\begin{figure}[t]
    \centering
    \includegraphics[width=0.90\columnwidth]{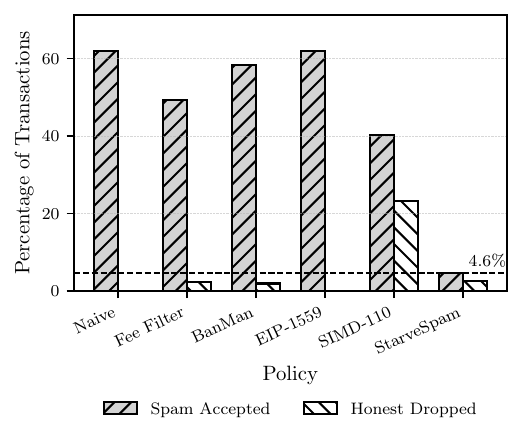}
    \caption{Percentage of transactions accepted or dropped by each policy.}
    \label{fig:policy_comparison}
\end{figure}

Fig.~\ref{fig:policy_comparison} shows the number of spam transactions accepted (false negatives) and honest transactions dropped (false positives) under each policy after $50{,}000$ transactions. \systemname{} achieves the best trade-off, accepting only $4.6\%$ of spam while dropping around $3\%$ of honest transactions. By comparison, Naive, Fee Filter, BanMan, and EIP-1559 all allow over 50\% of spam through. Although SIMD-110 significantly reduces spam acceptance, it drops a much larger share of honest traffic, demonstrating poor fairness. \systemname{} combines precise spam filtering with adaptive throttling to achieve the most optimal trade-off: it blocks the majority of spam while minimizing the loss of honest transactions.

\subsection{Reputation Evolution Over Time}
This experiment evaluates \systemname{}'s ability to dynamically adjust peer reputations based on observed behavior. The goal is to assess whether the system can distinguish honest from malicious nodes and support recovery for those that improve over time. \\
\indent We simulate a network of $100$ peers, divided into three behavioral categories: honest, Sybil, and reforming. All nodes start with equal reputation ($0.5$). Honest nodes consistently relay legitimate traffic, while Sybil nodes exhibit spam-like behavior throughout the simulation. Reforming nodes behave like Sybils initially but switch to honest behavior at time step $50$. Each peer's reputation is updated locally over $100$ time steps using the reputation mechanism described in Section~\ref{sec:system_design}.

\begin{figure}[t]
    \centering
    \includegraphics[width=0.86\columnwidth]{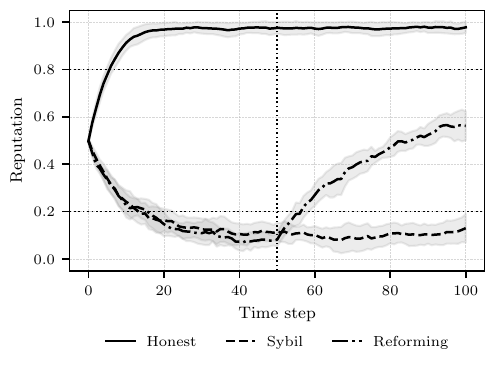}
    \vspace{1mm}
    \caption{Reputation trajectories for honest, Sybil, and reforming nodes.}
    \label{fig:rep_evolution}
\end{figure}

Fig.~\ref{fig:rep_evolution} shows the average reputation of each peer class over time. Honest nodes quickly converge above $0.8$, while Sybil nodes degrade below $0.2$. Reforming peers recover after switching behavior, demonstrating that \systemname{} supports adaptive trust without permanent penalties. Shaded regions indicate each group's standard deviation, capturing behavioral variance. These results confirm that \systemname{} distinguishes peer intent and maintains fairness in dynamic settings.

\subsection{Spam Propagation Suppression}

We evaluate how well each policy contains spam propagation across the network by simulating transaction relay over a randomized P2P overlay. Each transaction starts from a random node and propagates to neighbors unless filtered by local policies. We track and report the average number of nodes that each spam or honest transaction reaches under three strategies: Naive, BanMan, and \systemname{}.

We restrict our comparison to these three policies as they represent distinct and relevant points in the design space. Naive flooding models the absence of filtering, providing an upper bound on reachability. BanMan reflects Bitcoin's in-protocol peer discouragement mechanism, applying local penalties to misbehaving senders based on fixed rules. \systemname{}, in contrast, uses decentralized, adaptive reputation to enforce propagation limits. We omit global rate-limiting, centralized moderation, or application-layer filters, as these are incompatible with the decentralized, relay-layer setting we target.

\begin{figure}[t]
    \centering
    \includegraphics[width=0.90\columnwidth]{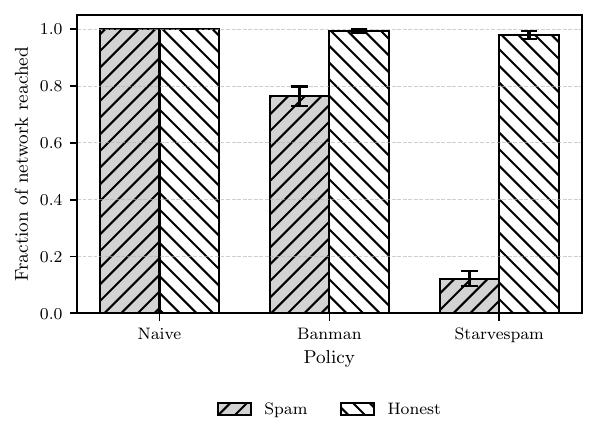}
    \caption{Average network coverage of spam and honest transactions under different relay policies.}
    \label{fig:gossip_propagation}
\end{figure}

Fig.~\ref{fig:gossip_propagation} shows the results. Under the Naive policy, both spam and honest transactions reach the entire network, as expected in unrestricted flooding. BanMan achieves modest suppression, blocking some spam at intermediate hops, but still allows broad propagation. In contrast, \systemname{} sharply reduces spam reach, with most spam transactions failing to spread beyond a small subset of peers. Meanwhile, honest transactions continue to propagate effectively, nearly matching the coverage of the Naive baseline.

\section{Conclusion}
Spam remains a persistent and costly challenge in blockchain networks, where permissionless access and cheap identities allow adversaries to degrade availability, inflate fees, and disrupt services. Existing defenses, primarily economic deterrence or protocol-specific heuristics, struggle to balance openness with resistance to abuse, often penalizing legitimate users or failing to suppress coordinated attacks.

We introduced \systemname{}, a decentralized, reputation-based spam mitigation system designed for the transaction relay layer. By leveraging local observations of peer behavior, \systemname{} enables nodes to adaptively throttle abusive senders without requiring global coordination, protocol changes, or centralized infrastructure. This addresses a key gap in existing stateless, identity-agnostic protocols. Because it is fully local and modular, \systemname{} can be incrementally adopted by relay nodes with minimal integration overhead.

Through simulations grounded in a real-world spam event, we demonstrated that \systemname{} achieves high accuracy in suppressing spam, preserves accessibility for honest users, and significantly limits the spread of malicious transactions across the network. Compared to fee-based or rule-based baselines, \systemname{} offers a more flexible and fair foundation for resilient spam prevention in decentralized environments.

Future work includes extending \systemname{} to multi-chain and Layer 2 architectures, integrating privacy-preserving identity systems, and validating its design in live testnet deployments. Further directions include, refining the scoring model with machine learning or game-theoretic incentive alignment, and studying deployment incentives to promote adoption across heterogeneous relayers.

\bibliographystyle{IEEEtran}
\bibliography{references}

\end{document}